\documentclass[aps,prc,twocolumn,showpacs,superscriptaddress,floatfix]{revtex4}

\usepackage{graphicx}
\usepackage{amsmath,amssymb}
\usepackage{times,txfonts}

\begin{document}

\title{Pygmy dipole response of proton rich argon beyond the random phase approximation}

\author{C. Barbieri}
\affiliation{Gesellschaft f\"ur
  Schwerionenforschung Darmstadt, Planckstr. 1, D-64259 Darmstadt,
  Germany}

\author{E. Caurier}
\affiliation{Institut de Recherches Subatomiques, Universit\'e Louis
  Pasteur, F-67037 Strasbourg, France}

\author{K. Langanke}
\affiliation{Gesellschaft f\"ur
  Schwerionenforschung Darmstadt, Planckstr. 1, D-64259 Darmstadt,
  Germany}
\affiliation{Institut f\"ur Kernphysik, Technische
  Universit\"at Darmstadt, Schlossgartenstr. 9, D-64289 Darmstadt,
  Germany}

\author{G. Mart\'inez-Pinedo}
\affiliation{Gesellschaft f\"ur
  Schwerionenforschung Darmstadt, Planckstr. 1, D-64259 Darmstadt,
  Germany}

\date{\today}

\begin{abstract}
The occurrence of a pygmy dipole resonance in proton rich ${}^{32,34}$Ar is studied using the unitary correlator operator method interaction V$_{UCOM}$, based on Argonne V18. Predictions from the random phase approximation (RPA) and the shell model in a no-core basis are compared.
 It is found that the inclusion of configuration mixing up to two-particle--two-holes broadens the pygmy strength slightly and reduces sensibly its strength, as compared to the RPA predictions. 
For ${}^{32}$Ar a clear peak associated with a pygmy resonance is found. For ${}^{34}$Ar, the pygmy states are obtained close to the giant dipole resonance and mix with it.
\end{abstract}

\maketitle


\section{Introduction}
\label{intro}

The recent advances in the experimental techniques for radioactive beams have fueled several studies of the properties of exotic nulcei, away from the line of beta stability. One of the most interesting results is the discovery of low-lying dipole strength in neutron rich isotopes which is interpreted as a pygmy dipole resonance (PDR). This new excitation mode is then explained as the resonant oscillation of the weakly bound neutron skin against the isospin saturated proton-neutron core. Typically, one observes in nuclei with a neutron excess, N$>$Z, a concentration of electric dipole states close to the particle emission treshold. These carry a small fraction of the Thomas-Reiche-Kuhn (TRK) sum rule, which increase with the charge asymmetry of the nucleus.
The PDR is also interesting because of its astrophysical implications. Although it carries only a small fraction of the total dipole strength, the occurrence of increased strength at the particle separation treshold can enhance radiative capture cross sections~\cite{Gor.02,Gor.04}, which can have strong effects on the $r$-process nucleosynthesis and on the abundance distribution of elements.
 In addition, the thickness of the nuclear skin is directly related to the density dependence of the symmetry energy. Recently, data on the PDR resonance have been used to constrain models for the symmetry energy~\cite{Pie.06}.

 The onset of low-lying E1 strength has been reported even in stable nuclei with moderate proton-neutron asymmetry such as $^{44,48}$Ca and $^{208}$Pb~\cite{Rye.02,End.03,Har.04}; see Ref.~\cite{Kne.06} for a review of high precision photon scattering experiments. For unstable nuclei with large neutron excess, a sizable fraction of low-lying E1  strength was obseved in $^{20,22}$O~\cite{Lei.01} and $^{\rm 130,132}$Sn~\cite{Adr.05}. 
Several theoretical models have been employed to study the nature of this low-energy dipole strength. Recent works employed the Skyrme Hartree-Fock (HF) plus quasiparticle random phase approximation (RPA) with phonon couplings~\cite{Col.01,Sar.04}, the quasiparticle phonon model~\cite{Tso.04,Tso.04b,Tso.07}, and the relativistic quasiparticle RPA (RQRPA)~\cite{Paa.03,Paa.05}.
In a recent work, the PDR was analysed including effects of particle-vibration coupling on top of the RPA approach~\cite{Lit.07}. The inclusion of low-lying phonons mainly correspond to consider explicit admixtures of two-particle--two-hole (2p2h) states and increase the fragmentation of the dipole distribution. In neutron rich isotopes it was found that this effect can generate a shift in the position of the PDR but otherwise does not noticeably change its characteristics.

On the proton rich side, nuclei with an excess of protons over neutrons are found only for Z$\leq$50. Due to the Coulomb repulsion, the proton drip line is much closer to the $\beta$ stability line and proton skins are possible only for the lightest isotopes. For these elements the multipole response is generally less collective. Although these facts seem to disfavor the existence of a proton PDR, a recent calculation suggested that this mode can actually be observed in medium-mass nuclei~\cite{Paa.05}. Using RQRPA calculations low lying pygmy states were obtained when approaching the proton drip line along the Ar isotopes and the N=20 isotones chains. 
However, no study of correlations beyond the RPA, such as in Ref.~\cite{Lit.07}, has been reported to date for the proton PDR.
In this work we will consider the proton rich isotopes $^{32,34}$Ar and compare predictions for the dipole strength as obtained in the RPA with those from shell model (SM) studies in a no-core cofiguration space. These approaches include complementary correlation effects since rather different portions of the Hilbert space are probed. Still it is preferable to perform such a comparison based on the same Hamiltonian. 

In the following, the unitary correlator operator method (UCOM)~\cite{Rot.04} will be employed to regularize the strong core of the realistic Argonne V18 potential~\cite{Wir.95}. The UCOM operator generates a unitary transformation within the many-body Hilbert space in which a weakly correlated wave function is mapped into one where strong short-range repulsion and tensor effects are explicitly manifest. If one applies this transformation to the Hamiltonian, rather than the wave function, it finds an expansion of the effective interaction into many-body terms. The $V_{UCOM}$ force is obtained by truncating the expansion at the two-body level. The result is an effective force that tames the strong short-range and tensor components of the original force and is therefore applicable to smaller basis sets. The subtle cancellation between large nuclear and kinetic contributions to the total energy is also accounted for in this expansion. Since the correlator operator is chosen in such a way that nucleon-nucleon phase shift are preserved, $V_{UCOM}$ can also be regarded as a realistic two-nucleon force on its own that is applicable, however, only to medium/low-energy processes. 
 Therefore, the UCOM method provides an interaction independent of the employed model space and can be applied meaningfully to both the RPA and the SM methods. At the same time it includes the correlation effects proper of modern high precision realistic interactions.

Details of the formalism employed in the calculations are discussed in Sec.~\ref{form}. The  RPA and the shell model results are compared in Sec.~\ref{sec:results}. Before presenting these results, we will discuss the dependence of our calculations on the parameters of the oscillator space, and use the Hartree-Fock (HF) approach to construct a basis to be used in the shell model studies, in Sec.~\ref{sec:Ar32conv}. Conclusions are drawn in Sec.~\ref{conl}.

\section{Formalism}
\label{form}

 Consistent RPA calculations were performed to guarantee the exact separation of the spurious center-of-mass motion. This means that our single particle basis was obtained by solving the HF problem and the same interaction was used for both HF and RPA.  To do this, one must take some care in treating the $d_{3/2}$ orbit which contains only two protons in the mean-field picture. In order to maintain consistency and to assure the separation of spurious states one must: (1) force uniform occupation for protons in this orbit when solving the HF equations, (2) account for its partial occupation  when solving the RPA problem, and (3) allowing for excitations of protons {\em both} from the $d_{3/2}$ to higher orbits {\em and} from lower orbits into the (half empty) $d_{3/2}$ level.
 This was done by considering the latter orbit both as a particle and as a hole state, with the respective depletion or occupation probabilities. Indicating the HF basis with {$a\equiv n_a l_a j_a \tau_a$} ($\tau$ is the isospin), the RPA eigenvalue equations for the amplitudes $X^J_{ab}$ and $Y^J_{ab}$ are
\begin{equation}
\label{eq:RPA}
\left(
\begin{array}{cc}
{\bf A^J} & {\bf B^J} \\
{\bf B^J}^{\ast} & {\bf A^J}^{\ast}
\end{array}
\right)
\left(
\begin{array}{c}
X^{J, \nu} \\
Y^{J, \nu}
\end{array}
\right) =\omega_{\nu}\left( 
\begin{array}{cc}
1 & 0 \\
0 & -1
\end{array}
\right)
\left( 
\begin{array}{c}
X^{J, \nu} \\
Y^{J, \nu}
\end{array}
\right)\; ,
\end{equation}
where
\begin{eqnarray}
A^J_{a b  ,  c d} &=& n_a \; \bar{n}_b \; H^J_{a b , c d} \; n_c \; \bar{n}_d 
 ~+~ \delta_{ac}\delta_{bd} (\varepsilon_{a} - \varepsilon_{b}) \; ,
\nonumber \\
B^J_{a b  ,  c d} &=& n_a \; \bar{n}_b \; H^J_{a b , d c} \; \bar{n}_d \; n_c \; ,
\label{eq:AB}
\end{eqnarray}
$J$ is the angular momentum of the excited state, $\varepsilon_{a}$ is the HF single particle energy of orbit $a$, and $H^J_{a b , c d}$ the Pandya transform of the residual two-body interaction. The numbers $(n_a)^2$ define the occupation of orbit $a$ in the HF wave function, while $(\bar{n}_a)^2$ measures the unoccupied space. Hence, for the Ar isotopes we have
\begin{equation}
\label{eq:partocc}
 (n_a)^2 = 1 - (\bar{n}_a)^2 = 
 \left\{
 \begin{array}{lll}
   1 &  ~ &  \hbox{for $a$ fully occupied} \\
   0 &  ~ &  \hbox{for $a$ empty} \\
   1/2 &  ~ &  \hbox{for } a = \pi 0d_{3/2} \\
 \end{array}
 \right.
\; .
\end{equation}
In Eqs.~(\ref{eq:RPA}) and~(\ref{eq:AB}), the products $n_a \bar{n}_b$  are restricted to 1p1h configurations only, except for the partially occupied orbit.
 It must be stressed that the above prescription arise naturally when deriving the HF+RPA scheme from propagator (or Green's function) theory~\cite{FetWal,DicVan}: following Baym and Kadanoff~\cite{Bay.61,Bay.62}, conserving RPA equations are derived consistently from the HF self-energy and propagator. The HF propagator, however, contains both particle and hole poles for each partially occupied orbit, such as the $d_{3/2}$ in our case. We checked that neglecting proton excitations from lower orbits to the $d_{3/2}$ level no longer allows for the {\em exact} separation of the spurious center-of-mass mode (although the breaking is small and has negligible effects on the remaining E1 strength). The RPA results reported in this paper are obtained in the fully conserving approach.
In all cases we consider the isovector dipole operator
\begin{equation}
\label{eq:Q}
 \hat{Q}^{T=1}_{1m} = \frac{N}{N+Z} \sum^Z_{p=1} r_p Y_{1m} ~ - ~ \frac{Z}{N+Z} \sum^N_{n=1} r_n Y_{1m} \; ,
\end{equation}
which is corrected for the center-of-mass displacement.
In all presentations below, the calculated B(E1) strength to each final state $f$,
\begin{equation}
B_{fi}(E1) =  \frac{1}{2 J_i + 1} \; {\langle f \| \hat{Q} \| i \rangle }  \; ,
\end{equation}
was folded with a Lorentzian of width $\Gamma$=1~MeV,
\begin{equation}
R(E_x) = \sum_f B_{fi}(E1) \frac{\Gamma  / 2 \pi}{(E_x-\omega_f)^2 + \Gamma^2/4} \; .
\label{eq:rex}
\end{equation}

For all the calculations we employ the $V_{UCOM}$ interaction and subtract the kinetic energy of the center-of-mass motion. The intrinsic Hamiltonian is then written in the form of a two-body interaction,
\begin{equation}
 H_{int} ~=~ T ~+~ V - T_{cm} ~=~ \sum_{i < j} \left\{
    \frac{({\bf p}_i - {\bf p}_j)^2}{2 \; A \; m_N} ~+~ V_{i,j}
    \right\} \; .
\label{eq:H}
\end{equation}
In all cases, a UCOM operator was used which corresponds to a correlation volume of \hbox{$I_\vartheta = 0.09$ fm$^3$} for the tensor force. This correlator was tuned to reproduce the binding energies of $^4$He and $^3$H without the need of additional three nucleon forces~\cite{Rot.05}. Subsequent calculations sugggested that binding energies are reproduced, in perturbation theory, throughout the nuclear chart without additional corrections to this UCOM operator~\cite{Rot.06,Bar.06b}.

An harmonic oscillator basis with length $b_{HO}$=1.8~fm was used for all the RPA calculations, except for $^{32}$Ar when indicated. This basis was truncated according to the number of major shells $2n+l\leq N_{max}$~(with $N_{max}$=0,1...).
Calculations based on $V_{UCOM}$ generally converge in model spaces of 10 to 20 shells if an harmonic oscillator basis is used.
Standard applications of the no-core shell model (NCSM)~\cite{NCSM} are based on using harmonic oscillator bases and truncating the model space in terms of the maximum number of cross shell excitations. The reason for doing so is to allow for an exact separation of the center-of-mass motion. As it will be shown in  Sec.~\ref{sec:Ar32conv}, a very large number of oscillator functions would be required for converging the excitation energy of the pygmy state.
 Obviously, this is beyond the capability of present day computers and one needs to resort to a more realistic single particle basis.
 Below, we will expand the Hamiltonian, Eq.~(\ref{eq:H}), over the lowest HF orbits that are obtained while solving the corresponding HF+RPA problem. Since we will only consider model spaces in which all nucleons are active, we refer to this scheme as NCSM.
The secular problem was diagonalized using the shell model code ANTOINE~\cite{ANT.99,ANTWEB}. 
The dipole response was derived as usual by first applying the operator~$\hat{Q}$ to the J$^\pi$=0$^+$ ground state of $^{32,34}$Ar, and then using the resulting 1$^-$ wave function as a pivot for successive Lanczos iterations.

\section{results}
\label{sec:results}

\subsection{Dependence on the model space}
\label{sec:Ar32conv}

\begin{figure}
\includegraphics[width=\columnwidth,clip=true]{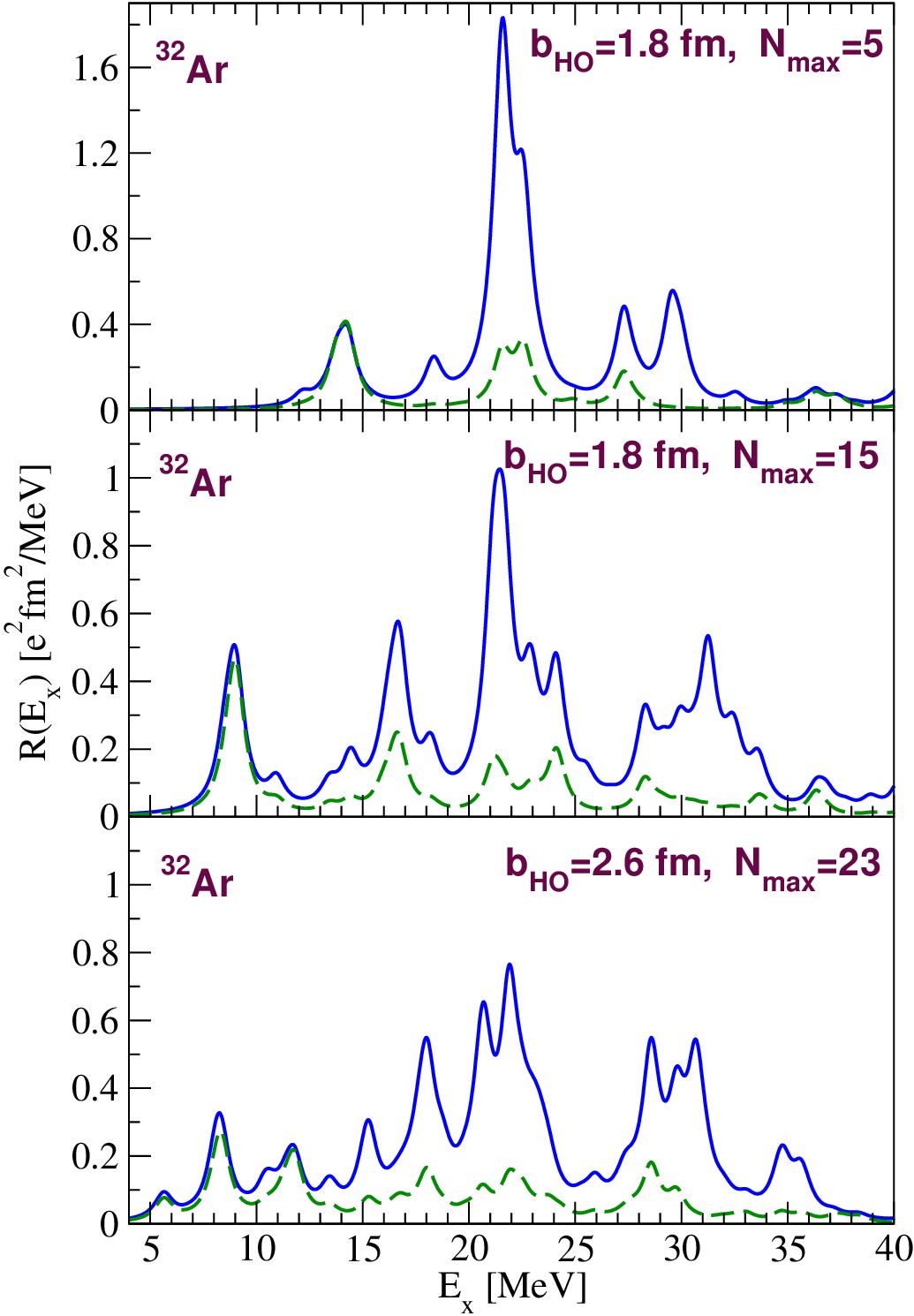}
\caption{(Color online) Isovector dipole strength of $^{\rm 32}$Ar obtained with HF+RPA, for different sizes of the models space ($N_{max}$) and harmonic oscillator lengths ($b_{HO}$).
The dashed line represents the contribution from protons only [first term in Eq.~(\ref{eq:Q})].}
\label{fig:rpa5-23}
\end{figure}

 Figure~\ref{fig:rpa5-23} shows the dipole strength distribution, Eq.~(\ref{eq:rex}), obtained with RPA for different model spaces and ocillator lengths. In all cases one can identify a peak on the low-energy tail of the isovector giant dipole resonance (IVGDR) which receives strength almost exclusively from proton excitations, that is from the first term on the r.h.s. of Eq.~(\ref{eq:Q}). As it will be discussed below, this represents the PDR.
 Unfortunately, very large configuration spaces are needed to converge the pygmy peak at low energies.
 This behavior is an artifact of employing an harmonic oscillator basis, which does not describe properly loosely bound states. 
 The dominant contributions to the PDR come from the collective excitation of protons to states just above the continuum threshold. Hence, rather diffuse wave functions may be needed to describe this strength properly. This implies large harmonic oscillator bases.
An inspection of the HF single particle energies of protons show that the orbits in the {\em sd} shell are well converged already for $N_{max}$=5, while the $pf$ orbits are found above the separation energy and keep changing when increasing the model space.

\begin{table}[t]
\begin{ruledtabular}
\begin{tabular}{lcccc}
          &  \multicolumn{2}{c}{E$_x \leq$ 12 MeV}
          &  \multicolumn{2}{c}{E$_x \leq$ 40 MeV} \\
\multicolumn{1}{c}{Basis} 
          & $\bar{\rm E}$ & $\sum$B(E1) & $\bar{\rm E}$ & $\sum$B(E1)\\
\hline
$b_{HO}$ = 1.8 fm: \\
$N_{max}$=5  &  -    &  -     & 23.06  & 7.13  \\
$N_{max}$=9  &10.71  & 1.03   & 22.69  & 8.01  \\
$N_{max}$=11 &10.12  & 1.08   & 22.14  & 7.86  \\
$N_{max}$=15 & 9.15  & 1.01   & 22.26  & 8.04  \\
$N_{max}$=19 & 8.72  & 0.97   & 22.14  & 8.05  \\
$N_{max}$=23 & 8.51  & 0.86   & 22.17  & 8.09  \\
\\
$b_{HO}$ = 2.6 fm: \\
$N_{max}$=15 & 9.03  & 1.13   & 21.99  & 8.27  \\
$N_{max}$=19 & 8.59  & 0.93   & 22.22  & 8.14  \\
$N_{max}$=23 & 9.28  & 1.14   & 22.12  & 8.09  \\
\end{tabular}
\end{ruledtabular}
 \caption[]{ Total E1 strength (in e$^2$fm$^2$) for $^{\rm 32}$Ar obtained with HF+RPA for energies below 12~MeV and 40~MeV. The centroids of these distributions are also reported (in MeV). }. 
\label{table}
\end{table}

In order to assess the sensitivity of our results to the choice of the model space, RPA calculations were performed  up to $N_{max}$=23 (24 major oscillator shells). 
For the largest model spaces, the HF ground state properties are independent of the oscillator length used. Hence, we have also preformed calculations with a larger oscillator length, $b_{HO}$=2.6~fm, to facilitates the description of states in the continuum.
The centroid and total strength of the resulting E1 distributions are reported in Tab.~\ref{table} for energies up to 12~MeV and 40~MeV.
It must be stressed that the pygmy behavior is reproduced by RPA theory even for the smallest model spaces considered here, and its existence is therefore a stable prediction. However, no strength is obtained in the low-energy region when a small number of oscillator functions is employed.
Increasing the space, the pygmy peak is lowered and it eventually stabilizes at around 9~MeV.
Tab.~\ref{table} shows that 16 major oscillator shells are required to converge the centroid of the IVGDR and the summed strength in the region below 12~MeV. Increasing the space further, small variations of the results still occur which are a result of discretizing the continuum with an increasing number of single particle orbits.
 The limit of an infinite space would be reached by solving directly the  continuum RPA equations. However, we do not expect that this would lead to sensible deviations for the integral quantities of Tab.~\ref{table}.

\begin{figure}
\includegraphics[width=\columnwidth,clip=true]{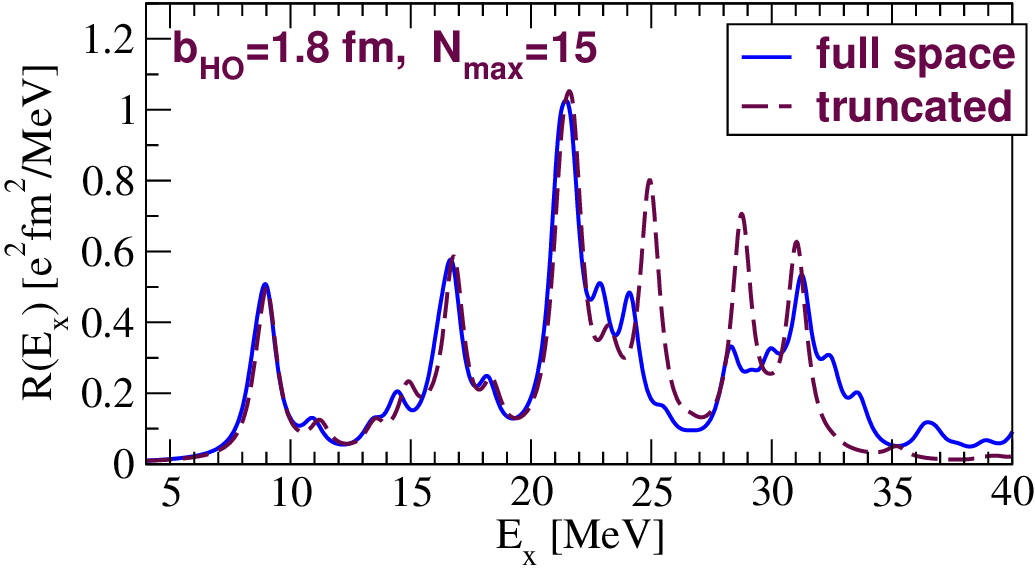}
\caption{(Color online) Dependence of the HF+RPA strength distribution on the truncation of the single particle basis. The same HF basis is employed in both calculations, and was derived in a $N_{max}$=15 harmonic oscillator space. Full line: all the HF states are used to solve the RPA equations. Dashed line: only the 6 lowest HF shells were retained.}
\label{fig:HFtrunc}
\end{figure}

 Although very large oscillator bases are needed to generate the proper HF wave functions, only the orbits near the Fermi surface are actually relevant for constructing the pygmy resonance. This is depicted in Fig.~\ref{fig:HFtrunc}, were the distribution obtained in the full $N_{max}$=15 model space is compared to the one obtained using the {\em same} HF basis but restricted to the 6 lowest shells. Solving the RPA equations in this basis is sufficient to reproduce the original results up to excitation energy of $\approx$~23~MeV .
The summed strength and centroid  below 40~MeV, obtained in this way, are 22.0~MeV and 7.97~e$^2$fm$^2$, which underestimates the total strength by only a percent due to truncation. Thus these quantities remain largely unaffected.
The truncation of the  HF basis reproduces the pygmy results obtained with the larger harmonic oscillator space. It is also small enough to allow for no-core shell model calculations, which are presented in Sec.~\ref{sec:Ar32ncsm}.

We note that not only the matrix elements of the Hamiltonian, Eq.~(\ref{eq:H}) were expanded in the new HF basis, but also the matrix elements of the dipole operator~$\hat{Q}$. Employing the correct one-body matrix elements of this operator was found to be crucial in order to reproduce the correct transition strengths.

\subsection{RPA results for $^{32}$Ar}
\label{sec:Ar32rpa}
  
\begin{figure}
\includegraphics[height=0.50\columnwidth,clip=true]{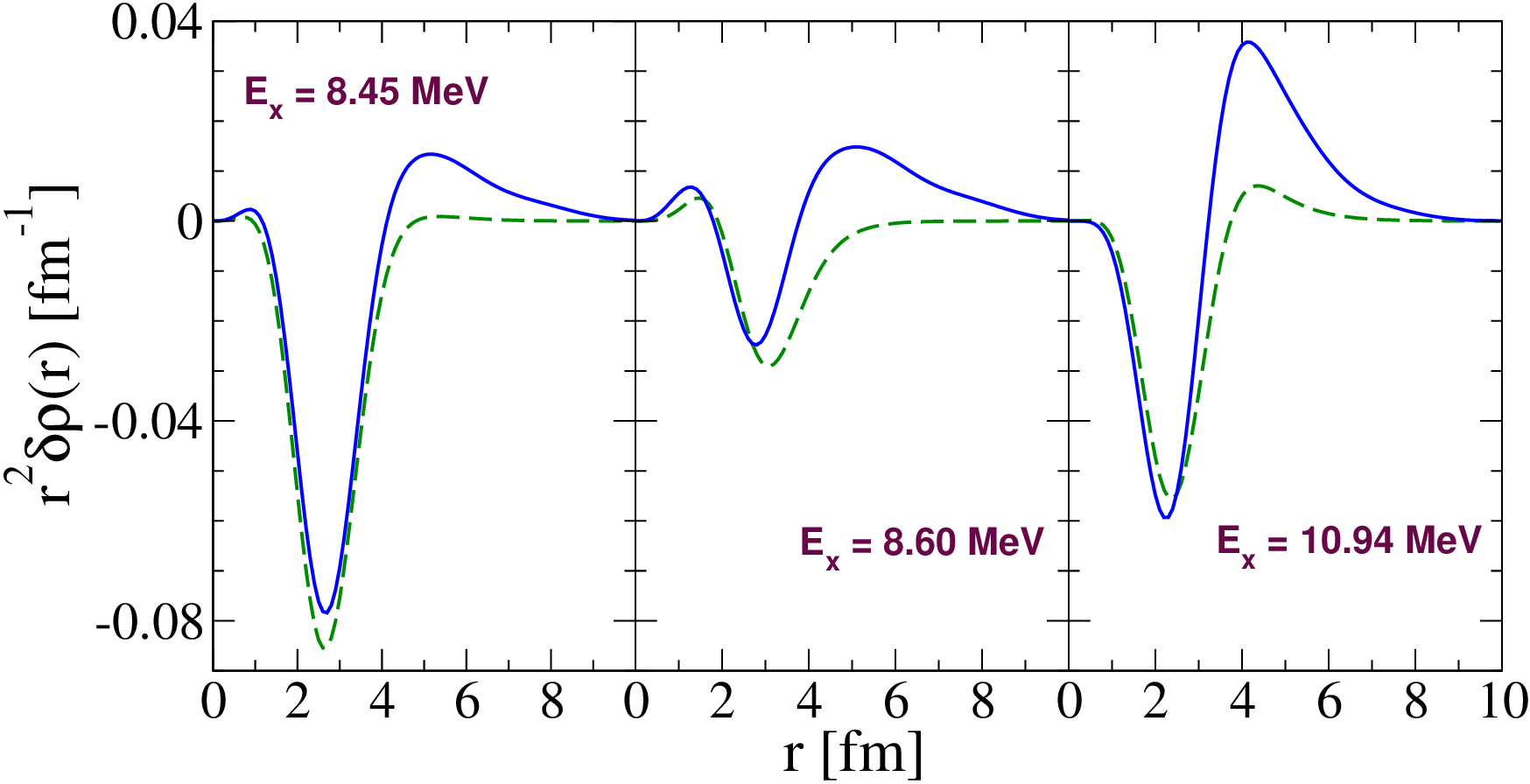}
\includegraphics[height=0.50\columnwidth,clip=true]{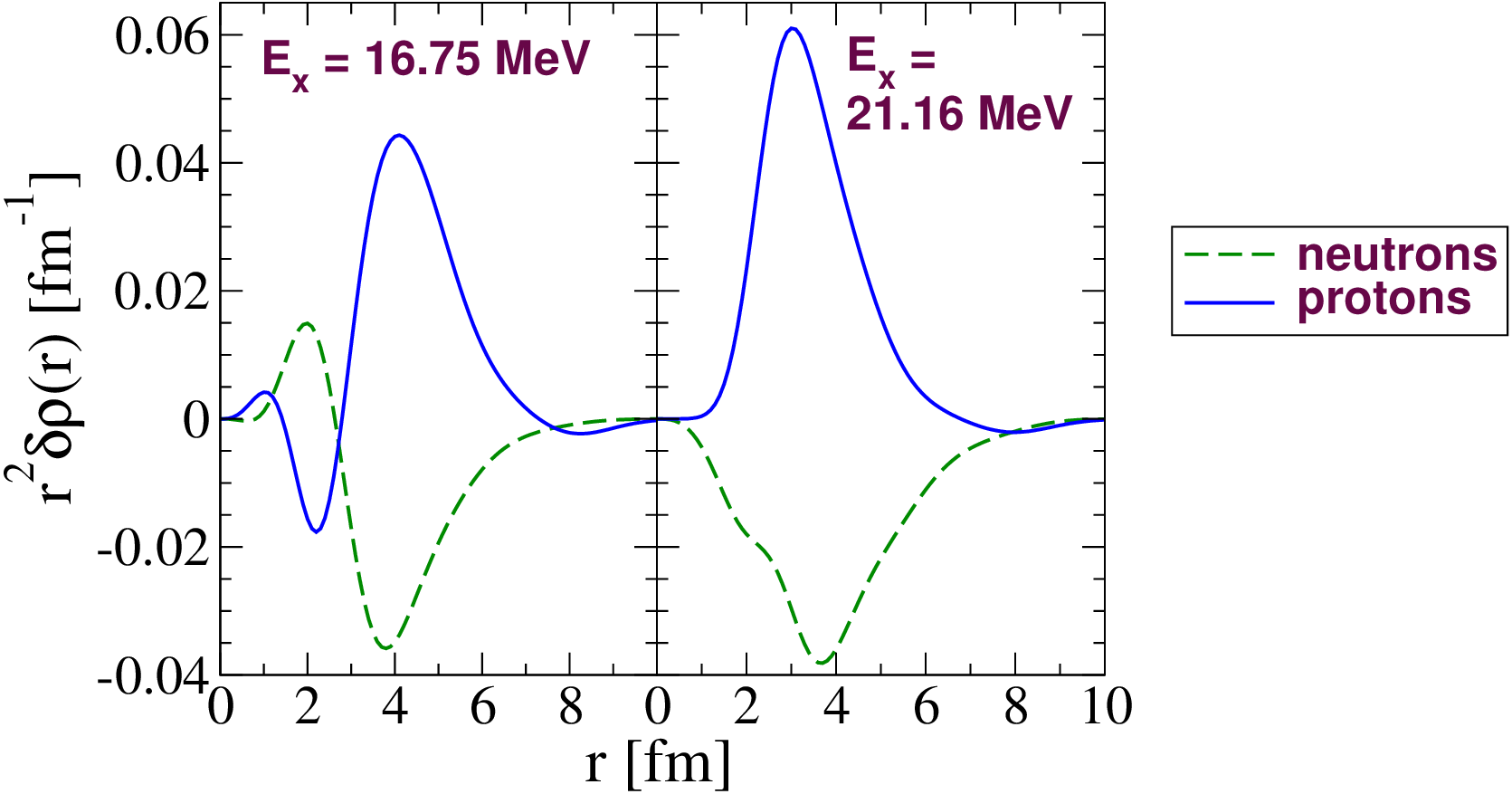}
\caption{(Color online) Proton and neutron transition densities for $^{32}$Ar obtained from RPA theory ($N_{max}$=15). Three states assocated with the proton pygmy resonance (at 8.45, 8.60 and 10.94~MeV) are compared to the IVGDR eigenstates at 16.75 and 21.16~MeV .}
\label{fig:Ar32tr-dens}
\end{figure}

From Tab.~\ref{table} we deduce that with, the present choice of the correlator (I$_\vartheta$=0.09fm$^3$) and at the RPA level, the V$_{UCOM}$ interaction predicts an isovector dipole strength of 8.15$\pm$0.10~fm$^2$e$^2$ within the first 40~MeV of excitation energy. The centroid of this distribution is E$_{IVGDR}$=22.15$\pm$0.15~MeV.
 This is close to the empirical estimate $E_{peak}$=$31.2/A^{1/3} + 20.6/A^{1/6}$=~21.4~MeV~\cite{Ber.75}. However, one should keep in mind that this formula was derived from data on stable nuclei with masses A$>$50, and it is just indicative in this case.
In the energy region up to 40~MeV the RPA calculation exhaust about 153\% of the TRK sum rule. This enhancement is expected for realistic interactions and it can be traced to the strong tensor component of the $V_{UCOM}$ force~\cite{Gaz.06}.
A similar enhancement has also been reported applying $V_{UCOM}$ to $^{\rm 4}$He~\cite{Bac.07}.
 Examples for the transition densities to states around 20~MeV are shown in Fig.~\ref{fig:Ar32tr-dens}. 
 These describe an out of phase oscillation of protons and neutrons and confirm the IVGDR nature of these excitations.

More interesting is the pygmy peak at the lower end of the GDR tail. As already noted, this receives E1 strength almost exclusively from protons---the first term on the r.h.s. of Eq.(\ref{eq:Q})---as shown by the dashed lines in Fig.~\ref{fig:rpa5-23}. The transition densities for the RPA states at 8.45, 8.60 and 10.94~MeV are reported in Fig.~\ref{fig:Ar32tr-dens} and show the typical behavior of the PDR: proton and neutrons move in phase in the nuclear interior while only protons are excited at the surface and extend ouside the nuclear core.
The total strength found below 12~MeV of excitation energy is 1.0$\pm$0.15~e$^2$fm$^2$, for $N_{max}\geq$15.
No strength is seen in this region if only 6 oscillator shells are retained. In this case the PDR is obtained at larger energies but it carries a similar strength (0.84~fm$^2$e$^2$ up to $E_x$=15~MeV).

\subsection{NCSM results for $^{32}$Ar}
\label{sec:Ar32ncsm}

\begin{figure}
\includegraphics[width=\columnwidth,clip=true]{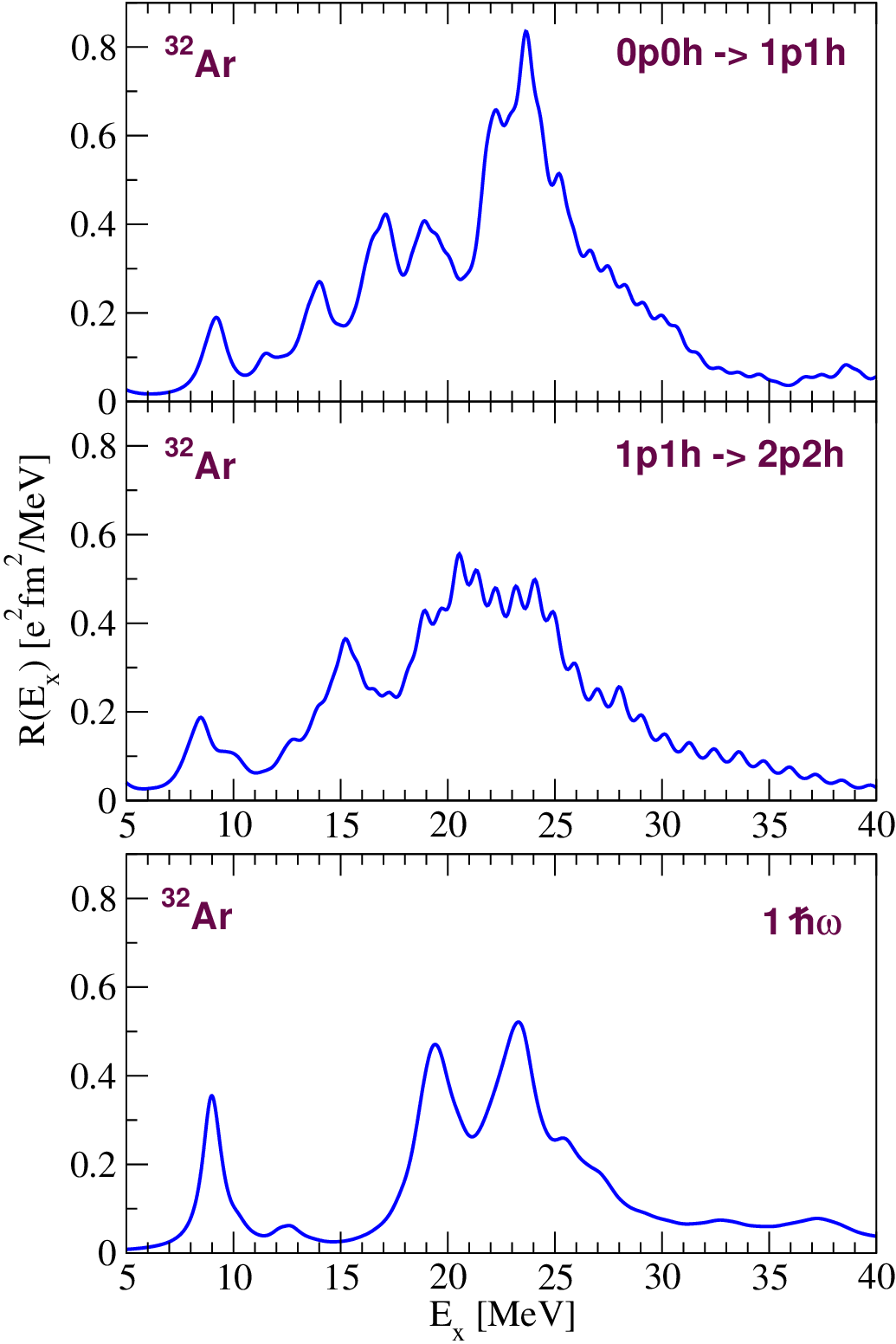}
\caption{(Color online) Top panels: isovector spectral strength obtained with the shell model for dipole excitations to the 1p1h and 2p2h configuration spaces. The first six HF orbits were included in the basis. Bottom panel: 1$\hbar\omega$ result, includig excitations up to the $pf$ shell only.}
\label{fig:Ar32ncsm}
\end{figure}

No-core shell model calculations were done in order to investigate the effect of correlations beyond RPA on the dipole distribution.
As it is apparent from Fig.~\ref{fig:rpa5-23} an oscillator basis $N_{max}$=15, corresponding to a 13$\hbar\omega$ model space, would be required to converge the energy of the PDR. The number of active single particle orbits can be reduced by employing the truncated HF basis  discussed in Sec.~\ref{sec:Ar32conv}. The proton HF orbits were used to expand the $V_{UCOM}$ interaction and the dipole operator.
 The explicit configurations employed were $(0s_{1/2}0p_{3/2}0p_{1/2}0d_{5/2})^{14-n_\nu}$$(1s_{1/2}0d_{3/2}0f_{7/2}\cdots0h_{11/2}0h_{9/2})^{n_\nu}$ for neutrons and  $(0s_{1/2}0p_{3/2}0p_{1/2}0d_{5/2}1s_{1/2}0d_{3/2})^{18-n_\pi}$ $(0f_{7/2}0f_{5/2}1p_{3/2}\cdots0h_{11/2}0h_{9/2})^{n_\pi}$ for protons. The model space was truncated in terms of the total number of particle-hole excitations ($n_\nu+n_\pi\leq n$). 
 This maintains all the particles active and allows for excitations from the core.
 The 0$^+$ ground state was first derived using a \hbox{$n$-particle--$n$-hole} (npnh) model space. Once this is done, the dipole operator, Eq.~(\ref{eq:Q}), connects this to states in a space containing up to (n+1)p-(n+1)h. By performing Lanczos diagonalization of the 1$^-$ final states in the larger configuration space one is guarantied to exhaust the total dipole strength. Note that for a 0p0h space, the neutrons are constrained to the lowest HF orbits while some mixing is possible 
 for protons, that can be excited from the lower shells into the half empty $d_{3/2}$ orbit.
 This already leads to a configuration space beyond the corresponding RPA theory.

The upper panels of Fig.~\ref{fig:Ar32ncsm} show the strength distributions obtained from the 0p0h-1p1h and 1p1h-2p2h configuration spaces. The same structure of a IVGDR resonance and pygmy peak is found, like for the RPA (Fig.~\ref{fig:HFtrunc}), but both the IVGDR distribution and PDR are broadenend by shell model correlations.
For the 2p2h model space, the IVGDR is centered at 21.66~MeV. Integrating up to 40~MeV excitation energy, one finds that the total strength is lowered by configuration mixing  and  exhausts 137\% of the TRK sum rule.
The response in the pygmy region is compared to the RPA approximation in Tab.~\ref{tab:pygmy}. Shell model correlations do not significantly alter the position of the PDR. On the other hand, its strength is reduced by a factor of two.

\begin{table}[t]
\begin{ruledtabular}
\begin{tabular}{lcccc}
          &  \multicolumn{2}{c}{$^{32}$Ar}
          &  \multicolumn{2}{c}{$^{34}$Ar} \\ 
          & $\bar{\rm E}_{pyg}$ & $\sum$B$_{pyg}$(E1)  & $\bar{\rm E}_{pyg}$ & $\sum$B$_{pyg}$(E1) \\
\hline
\\
RPA    &    1.0  &  9.15   &  0.8  &  12.7  \\
\\
1p1h   &  0.44   & 9.66    &  0.65 &  12.8  \\
\\
2p2h   &  0.49   & 8.95    &  0.62 &  11.6 \\
\end{tabular}
\end{ruledtabular}
 \caption[]{ Total dipole strength (e$^2$fm$^2$) and its centroind (MeV) in the region of the Pygmy resonace, as obtained in RPA and shell model with 1p1h and 2p2h configurations. The dipole dirtsibution was integrated up to 12~MeV of excitation energy for $^{32}$Ar and 14~MeV for $^{34}$Ar. }. 
\label{tab:pygmy}
\end{table}

The configuration spaces used above restrict most of the neutrons in the $d_{5/2}$ orbit and below. One may question whether neutron correlations in the $sd$ shell affect the properties of the PDR. To check this we performed a NCSM calculation in the 1$\hbar\omega$ model space, that is opening the $sd$ shell and allowing for one excitation across major shells. The strength distribution, plotted in the lower panel of Fig.~\ref{fig:Ar32ncsm}, shows no qualitative distortion of the pygmy peak. Hence, the approximation made restricting neutrons in the 0p0h configuation should not affects our conclusions noticeably.
 The PDR strength is also reduced here to about 60\% of the RPA value. However, at 1$\hbar\omega$ only excitation to the $pf$ shell are possible. Since this corresponds to a more severe truncation of the HF basis, we do not attempt to extract quantitative information on the PDR from this calculation.

\begin{figure}
\includegraphics[width=\columnwidth,clip=true]{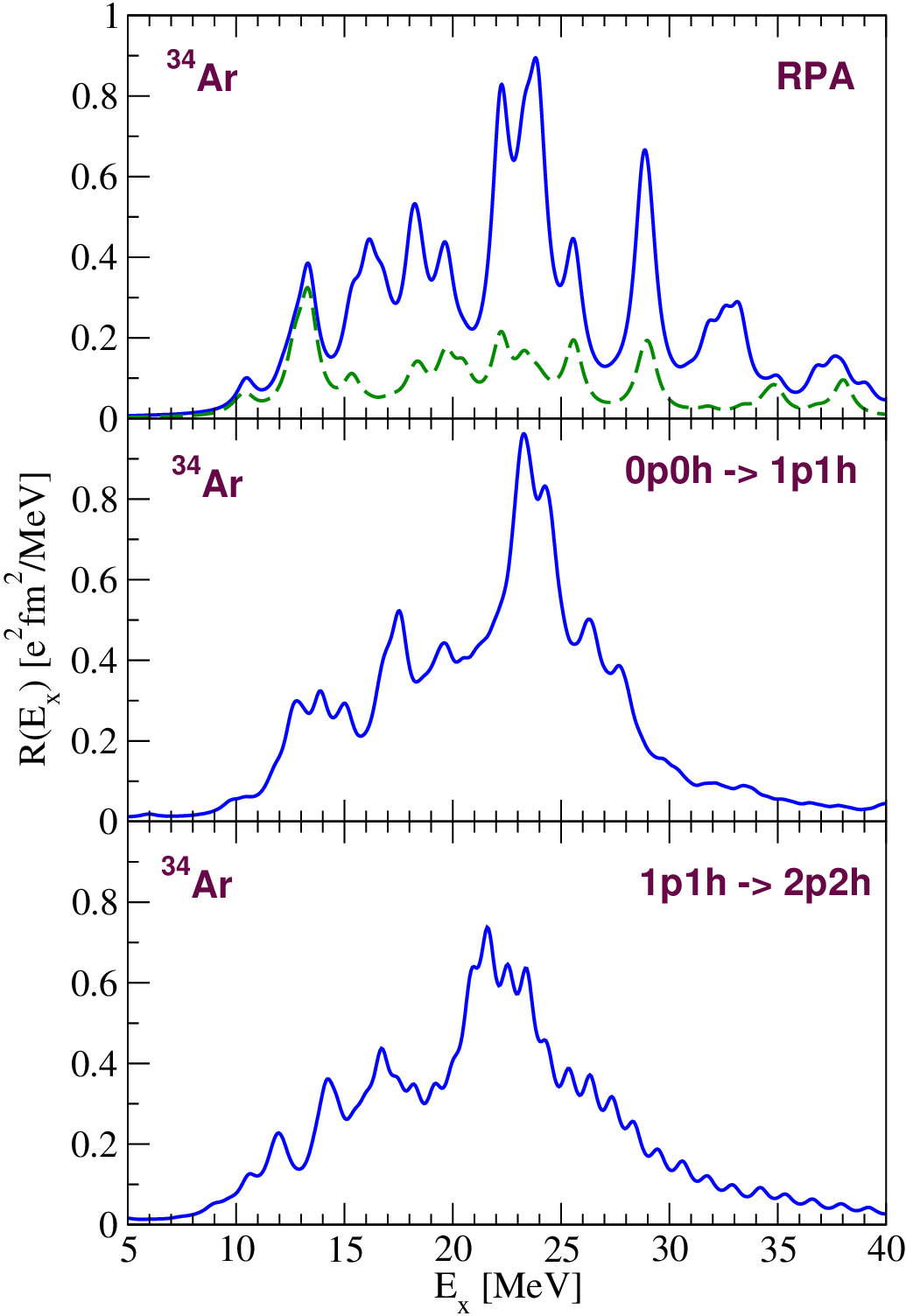}
\caption{(Color online) Isovector spectral strength for $^{34}Ar$ obtained with HF+RPA (top) and the shell model up to 1p1h and 2p2h (center and bottom). The full $N_{max}$=15 harmonic oscillator space was employed in the RPA calculations. The same HF basis derived for the RPA calculation is used for the shell model, but truncated to the lowest 6 shells. }
\label{fig:Ar34be1}
\end{figure}

\begin{figure}
\includegraphics[width=\columnwidth,clip=true]{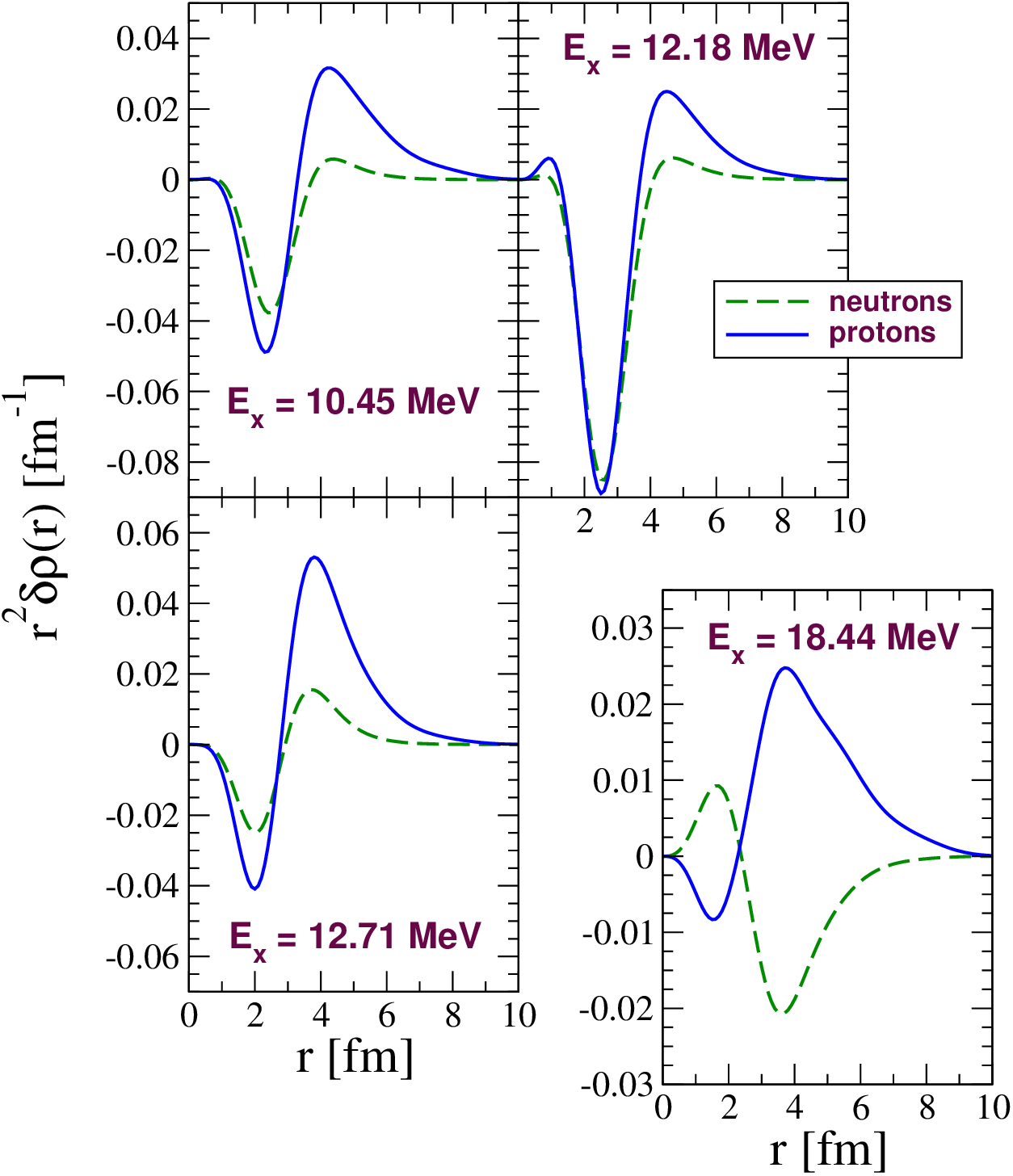}
\caption{(Color online) RPA transitions densities for the exitations to three pygmy states of $^{34}Ar$ (at 10.45, 12.18 and 12.71~MeV) and to a IVGDR state at 18.44~MeV.}
\label{fig:Ar34tr-dens}
\end{figure}

\subsection{Results for $^{34}$Ar}
\label{sec:Ar34}

 A similar analysis was carried out for $^{\rm 34}$Ar. The strength predicted from RPA theory is depicted in Fig.~\ref{fig:Ar34be1} for $N_{max}$=15.
 The centroid of the dipole distribution in the interval [0,40]~MeV is obtained as 23.3~MeV. Summing this strength within the same interval gives 8.69~e$^2$fm$^2$, which corresponds to 161\% of the TRK sum rule.
  These predictions do not vary by increasing the model space and the oscillator parameter. A similar analysis as done for $^{\rm 32}$Ar, suggests an uncertainty of $\pm$0.8~MeV and $\pm$0.2e$^2$fm$^2$ for these quantities.
  
 A pygmy resonance is also found for this isotope but closer to the centroid of the IVGDR. The dashed line in Fig.~\ref{fig:Ar34be1} shows that sizable proton contribution to the dipole response strength is found up to about 14~MeV. An analysis of the transition densities (Fig.~\ref{fig:Ar34tr-dens}) confirms that these states have the character of a pygmy mode. Instead, the RPA eigenstates found just a few MeVs above contribute to the lower tail of the IVGDR.
To estimate the total strength associated with the pygmy resonance, we integrated $R(E_x)$ up to 14~MeV and found 0.80~e$^2$fm$^2$.

No-core shell model calculations were performed in a model model space analogous to the $^{32}$Ar case, obtained by solving the HF equation for $^{\rm 34}$Ar in the same harmonic oscillator space as used for RPA. This basis was then truncated to 6 major shells and, at the 0p0h level, neutrons were constrained in the $d_{5/2}$ and $s_{1/2}$ orbits.
Also in this case we found a slight reduction of the total strength w.r.t. to the RPA result, yelding 8.21~e$^2$fm$^2$  for the 1p1h-2p2h configurations space. In this case the centroid of the IVGDR was lowered to 22.0~MeV and the energy weighted sum rule reduces to 143\% of the TRK value, in the first 40~MeV.
The strength resulting from the shell model calculation is plotted in the lower panels of Fig.~\ref{fig:Ar34be1} and shows a broad distribution of the E1 strength. Since for this nucleus the pygmy peak is obtained close to the giant resonance, the two are strongly mixed and one finds just a long tail spreading in the low-energy region. From the figure it is not possible to clearly identify a separate peak corresponding to the pygmy resonance.
Table~\ref{tab:pygmy} lists the summed strength below 14~MeV, where one expects to find most of the pygmy states. 
Contrary to the $^{32}$Ar case, the effects of configuration mixing on the low-energy dipole strength is moderate. A reduction of about 20\% w.r.t to the RPA result was found.

\section{Conclusions and discussion}
\label{conl}

The properties of the pygmy dipole resonance  in the proton rich isotopes $^{32}$Ar and $^{34}$Ar were investigated by comparing the predictions of the RPA and the shell model in a no-core configuration space. The V$_{UCOM}$ interaction was employed in all cases as a low-energy realistic Hamiltonian. In this approach the nuclear force is tamed to account for the effects on short range correlations and can be directly applied in large scale RPA and shell model calculations.
 In general a large number of harmonic oscillator shells is still required to converge the distribution of dipole strength in  the low-energy region. However, it was seen that the number of active orbitals can be significantly reduced using the HF single particle basis. Therefore, comparisons could be done employing the same Hartree-Fock basis for the RPA and shell model.

 For $^{32}$Ar, enhanced low-lying strength was found at energies up to 12~MeV, which could be identified as proton PDR. The corresponding strength distribution is peaked at about 9~MeV, considerably above the experimental proton separation energy of 2.4~MeV~\cite{toi.96}. This should not come as a surprise, since oscillations of the proton skin need to overcome the Coulomb barrier.
  The RPA approximation predicts about 1.0~e$^2$fm$^2$ in this energy region. Additional correlations, as accounted for in shell model studies, reduce this strength. Furthermore the pygmy peak is slightly broadened due to a larger number of configurations. By explicitely including up to 2p2h states, the pygmy strength is halved to 0.49~e$^2$fm$^2$.
The results obtained here predict a well defined pygmy peak separated from the IVGDR, in accordance with the previous RQRPA calculations of Paar et al.~\cite{Paa.05}.
 Excitations associated with a PDR were also found for $^{34}$Ar, which has only two more protons than neutrons. The present calculations place its strength at energies up to 14~MeV and close to the IVGDR. The total strength found up to this energy is also reduced by shell model calculations, as compared to RPA.  However, the mixing with the nearby giant resonance states leads to a rather uniform response without a well defined pygmy peak.

 We note that the neutron PDR observed in larger nuclei is typically found below 10~MeV and its excitation energy lowers with increasing mass.
 On the other hand, little is still known for medium-mass nuclei. Unstable oxygen isotopes, for example, show a behavior more similar to the NCSM results of Fig.~\ref{fig:Ar34be1}, i.e. a broad dipole response spreading continuously from the giant resonance region down to the PDR~\cite{Lei.01}. The proton PDR, if discovered, could also have a different qualitative behavior, as the Coulomb repulsion affects its excitation energy.
 The shell model predicts a centroid of the PDR of $\approx$11.6~MeV  for $^{\rm 34}$Ar (see Tab.~\ref{tab:pygmy}) which may appear particularly large. However, this could be understood in view of the expected Coulomb effect. The large centroid might also be due to a shortcoming of the V$_{UCOM}$ interaction: calculations based on this force have reported too small radii~\cite{Rot.06} and correspondingly they overestimate the energy of GDRs~\cite{Paa.06}, at the RPA level. Although, recent second RPA calculations which include 2p2h configurations (as done here) seem to cure this issue~\cite{Pap.07}. The inclusion of three-nucleon forces may also play a role.
Planned experiments, aimed at observing low-energy dipole response of proton rich Ar isotopes~\cite{Bor.06}, will help testing the accuracy of the present approach.
If the pygmy states in $^{\rm 34}$Ar are actually observed within a few MeVs of the IVGDR tail, a substantial mixing should be expected. This would result in a broad PDR, seen as an extended low-energy tail of the isovector dipole distribution.

\vspace{0.5cm}

\acknowledgments
We thank K. Boretzky, T. Aumann, R. Roth, and S. Bacca for several useful discussions.


\end{document}